\documentclass[twocolumn,showpacs,preprintnumbers,amsmath,amssymb,prl]{revtex4}
\pdfoutput=1

\usepackage{graphicx}
\usepackage{dcolumn}
\usepackage{bm}
\usepackage{subfigure}
\usepackage{amssymb}

\begin{document}

\preprint{APS/123-QED}

\title{Lateral magnetic anisotropy superlattice out of a single (Ga,Mn)As layer}

\author{R.G. Dengel, C. Gould, J. Wenisch, K. Brunner, G. Schmidt and L.W. Molenkamp}
\affiliation{%
Physikalisches Institut (EP3), Universit\"at W\"urzburg, Am Hubland.
D-97074 W\"urzburg, Germany
}%


\date{\today}

\begin{abstract}
We use lithographically induced strain relaxation to periodically
modulate the magnetic anisotropy in a single (Ga,Mn)As layer. This
results in a lateral magnetoresistance device where two non-volatile
magnetic states exist at zero external magnetic field with
resistances resulting from the orientation of two lithographically
defined regions in a single and contiguous layer.

\end{abstract}

\maketitle


Strain control of the magnetic anisotropy in (Ga,Mn)As is currently
a focus of spintronics research. Recent reports include lithographic
strain control \cite{Humpfner2007,Wenisch2007} and related devices
\cite{pappertNatPhys,Wunderlich2007,RanieriCondmat}, as well as the
use of piezoelectric elements to electrically control the imposed
strain \cite{UnottPiezoCondmat,FurdynaPizoCondmat}. So far, all of
these devices imposed an homogeneous anisotropy onto the targeted
structure. In this paper, we show that this process can be taken
much further and demonstrate the fabrication and characterization of
a lateral anisotropy superlattice. Our method involves patterning of
the surface of a ferromagnetic semiconductor layer in order to allow
partial strain relaxation. This is achieved by lithographically
defining nanometer scale stripes into a (Ga,Mn)As layer and
partially etching into the ferromagnetic layer. The unetched part of
the layer remains pseudomorphically strained with its magnetic
anisotropy relatively unchanged, while the stripes anisotropically
relax and gain a strongly uniaxial magnetic anisotropy. By varying
the depth of etching, we can control the ratio of uniaxial to
biaxial anisotropy material. Our investigations reveal that for a
specific ratio, the resulting structure has alternating regions of
uniaxial and biaxial anisotropy, and thus constitutes a anisotropy
superlattice structure.



\begin{figure}[h]
\includegraphics[width = 1.0\columnwidth]{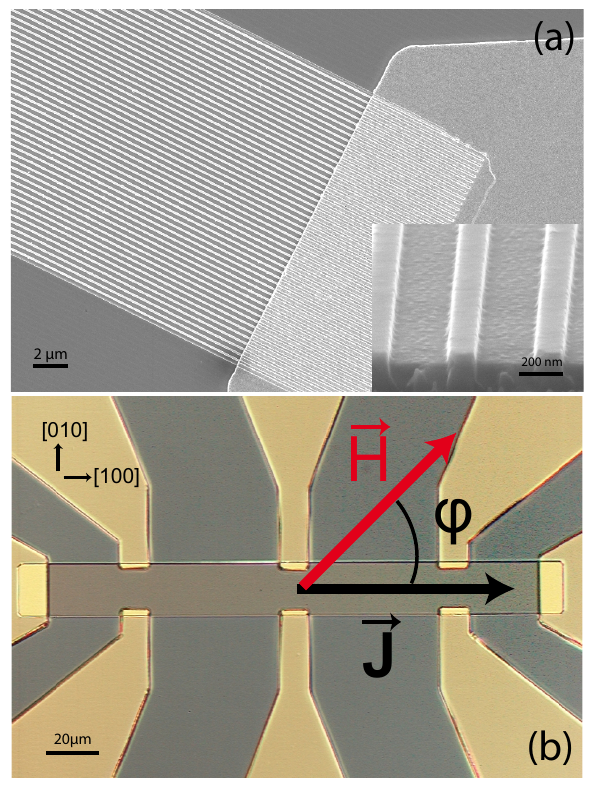}
\caption{\label{fig:sample} a) SEM image of the mesa and the stripes
after patterning. Inset: high magnification cross sectional view of
a few stripes. b) Picture of the Hallbar geometry relative to the
crystal direction with a sketch of the current and the applied
magnetic field.} \vspace{-0.5cm}
\end{figure}

The principle of our method is based on the nature of the
ferromagnetism in (Ga,Mn)As. The magnetic interaction between the
randomly distributed dilute Mn moments is mediated by itinerant
holes through Zener double exchange. Because of spin orbit coupling,
the wavefunction of these holes takes on anisotropies which reflect
the symmetry of the lattice. When (Ga,Mn)As is grown on GaAs (001),
it is compressively strained, and for the doping concentration used
in our transport samples, at 4K, the primary anisotropy is an
in-plane biaxial anisotropy along the [100] and [010] crystal
directions \cite{Sawicki2004}. When lithographic techniques are used
to trigger anisotropic deformation of the lattice along a given
direction, this change in symmetry is reflected in the hole
wavefunctions, and thus leads to a modification of the magnetic
anisotropy.

An important distinction between the anisotropy in dilute magnetic
semiconductors and traditional metals is worth noting. In metals, a
common way to achieve anisotropy engineering is through shape
anisotropy. However, because the strength of shape anisotropy scales
with saturation magnetization, it is a relatively weak effect in
(Ga,Mn)As and cannot effectively compete against the crystalline
anisotropy terms. For example, the uniaxial shape anisotropy field
expected \cite{Aharoni1998} from the 38 nm deep etched stripes used
in this study is $<6$ mT, much to small to compete with the
crystalline anisotropy of $\sim 100$ mT.

The samples are made from a typical  70 nm
Ga$_{0.965}$Mn$_{0.035}$As layer \cite{Giselagrowth} with a Curie
temperature of 60 K grown on a semi-insulating (001) GaAs buffer and
substrate. An array of 500 stripes of 150 nm width by 200 $\mu$m
length, with a period of 400 nm, and aligned along the [100]
crystallographic direction, is defined by e-beam lithography into a
positive bilayer resist, followed by deposition of 15 nm Ti and
lift-off. This pattern is transferred into the 70 nm (Ga,Mn)As layer
by chemical assisted ion beam etching (CAIBE) using the Ti stripes
as a mask. By properly selecting the etching depth, we fabricate
samples with stripes from $\sim 20$ to $\sim 60$ nm in thickness,
leaving a continuous layer from 50 to 10 nm between the stripes. The
Ti is then removed by a dip into HF:H$_{2}$O (1:200). A mesa of 20
$\mu$m $\times$ 200 $\mu$m, is defined onto the center of the array,
parallel to and containing $\sim 50$ stripes, by optical lithography
using a positive resist. Chemical wet etching is used to remove
material around this mesa to a depth of $\sim 60$ nm into the GaAs
substrate. A second optical lithographic step is then used to define
Au/Ti contact pads for electrical access. Fig.~\ref{fig:sample}a
presents a scanning electron microscope (SEM) image of one of the
ends of the 20 $\mu$m wide mesa containing the stripes, where they
meet the Au contact. The inset shows a close up of some of the
stripes, while Fig.~\ref{fig:sample}b shows the final sample layout.
The resulting thickness of the stripes is determined from SEM images
taken at tilted angles as well as by a stylus profiler system
(Dektak 6M).

The samples are characterized by magnetoresistance studies in a
magnetocryostat fitted with a vector field magnet capable of
applying fields of up to 300 mT in any direction. For our purposes,
the magnetic field will be confined to the sample plane, and its
direction $\varphi = 0$ is defined with respect to the current
direction flowing along the stripes. Given that the contact
resistance ($<10~\Omega$) is negligible compared to the resistance
of the stripes , a simple two terminal measurement configuration is
chosen in order to maximize our sensitivity to the distribution of
all the stripes instead of preferentially measuring those near the
edges. We apply a constant voltage of 10 mV to the end leads of the
Hall-bar and measure the resulting current $J$ in order to determine
the resistance while the magnetic field is swept in the sample
plane. In ferromagnetic materials the longitudinal resistance of a
sample is strongly dependent on the angle between the magnetization
$\vec{M}$ and the current $\vec{J}$. In (Ga,Mn)As, this anisotropic
magnetoresistance (AMR) \cite{Jan1957, MCGUIRE1975} has a resistance
minimum when $\vec{M} \parallel \vec{J}$ and a maximum when $\vec{M}
\bot \vec{J}$ \cite{Baxter2002}.

\begin{figure}
    \subfigure{\label{fig:20nm}}
    \subfigure{\label{fig:60nm}}
    \includegraphics[width = 1.0\columnwidth]{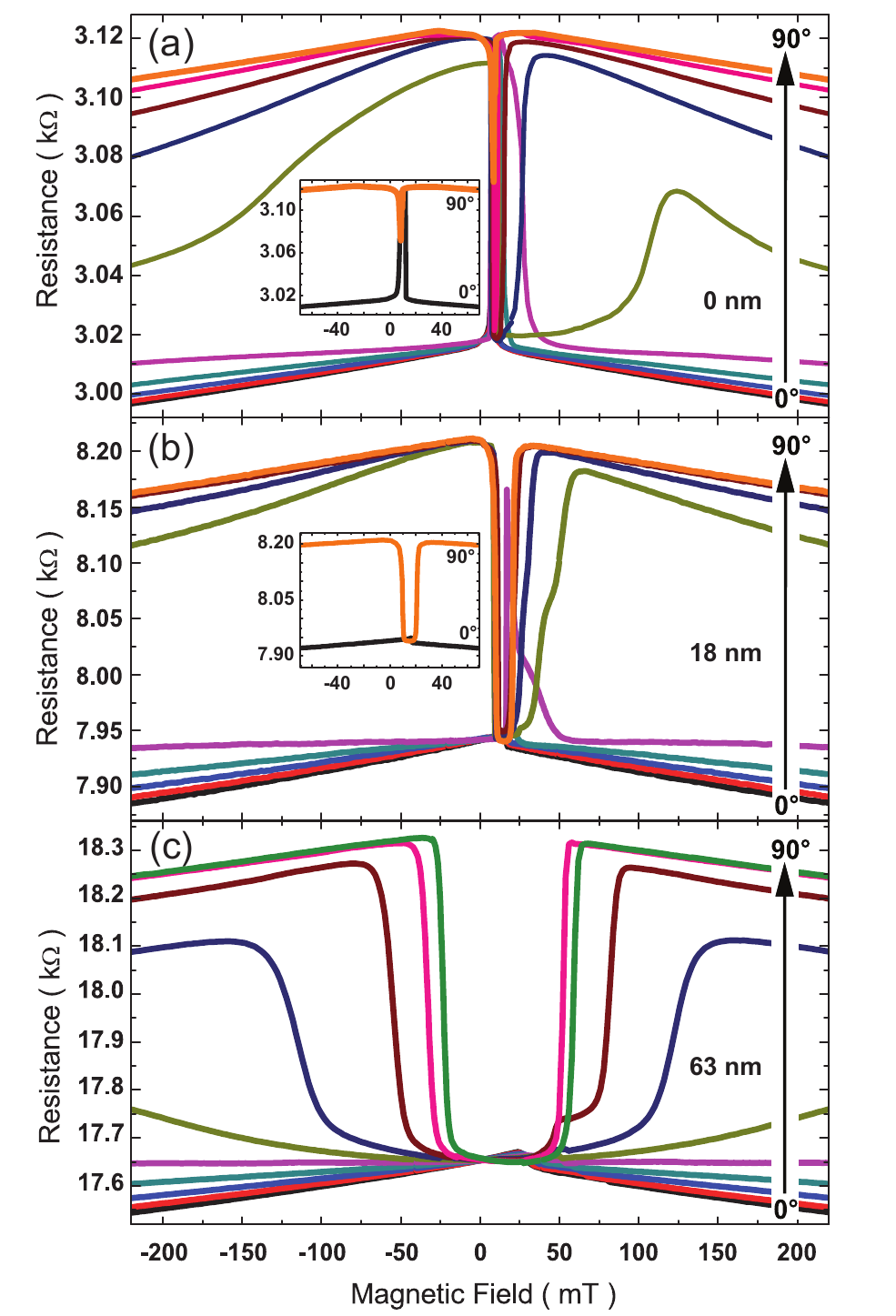}
\caption{MR scans at 4.2 K of patterned (Ga,Mn)As for various angles
$\varphi$ in steps of $10^{\circ}$ between magnetic field and
current. The field is swept from -300 mT to 300 mT. a) a Hall bar
with no etching, b) A sample with 18 nm thick stripes and biaxial
behavior; c) sample with 63 nm thick stripes which is fully
uniaxial. The insets in a) and b) show a zoom of the low field
region.} \vspace{-0.5cm}
    \label{fig:20nm60nm}
\end{figure}

We first consider measurements on a sample with stripes of 18 nm
thickness (Fig~\ref{fig:20nm60nm}b). In all curves, the magnetic
field is swept from -300 to 300 mT. For high magnetic fields the
magnetization $\vec{M}$ of the sample follows the applied external
magnetic field $\vec{H}$ and as expected, at $\pm$200 mT, the sample
has a maximum resistance for fields along 90$^{\circ}$,
perpendicular to $J$, and a minimum resistance for $H \parallel J$.
When $\vec{H}$ decreases, the magnetization $\vec{M}$ rotates
towards the nearest available easy axis. For this sample, we observe
that for field directions closer to 90$^{\circ}$, the result of this
relaxation is a high resistance state, whereas for field directions
nearer to 0$^{\circ}$ the sample relaxes to a low resistance state.
At $H = 0$ mT two stable states exist corresponding to the typical
biaxial behavior of (Ga,Mn)As. After $H=0$ mT in all curves,
magnetization reversal takes place through the nucleation and
propagation of two subsequent 90$^{\circ}$ domain walls resulting in
two distinct switching events. For reference, magnetoresistance
curves on the unetched layer are shown in Fig. 2a. In such a layer,
the low field features in the 0$^{\circ}$ and 90$^{\circ}$ curves
are of similar width, reflecting the near symmetry of these two
directions. In the layer with the 18 nm deep etched stripes, the
much wider feature in the 90$^{\circ}$ curve and the near absence of
any feature in the 0$^{\circ}$ curve, reveals a symmetry breaking
between the two directions, highlighting a uniaxial easy anisotropy
component along [100], the direction of the patterned stripes. The
sample thus shows primary biaxial behavior with the uniaxial
component imposed by the stripes playing only a secondary role.

The situation is reversed in Fig. 2c for the sample with stripe
thickness of about 63 nm. While this sample's behavior is similar to
the first at high fields, at $\vec{H} = 0$ mT all curves converge to
a single low resistance value corresponding to $\vec{M}
\parallel \vec{J}$. The sample is now totally dominated by the
uniaxial character of the stripes, and behaves as a simple uniaxial
magnet.

\begin{figure}
    \subfigure{\label{fig:40nm}}
    \includegraphics[width = 1.0\columnwidth]{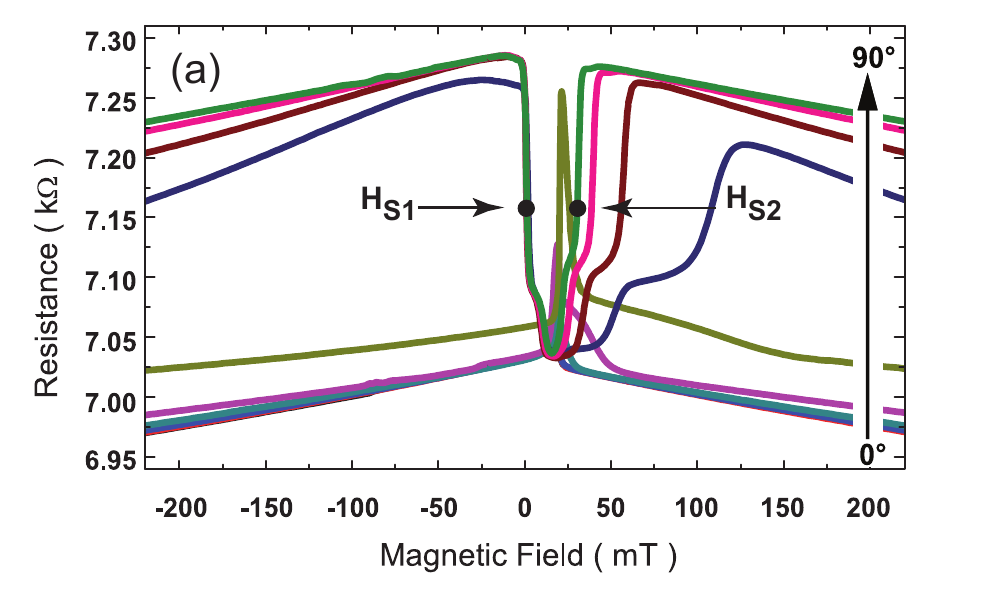}
    \subfigure{\label{fig:40nm_minors}}
    \includegraphics[width = 1.0\columnwidth]{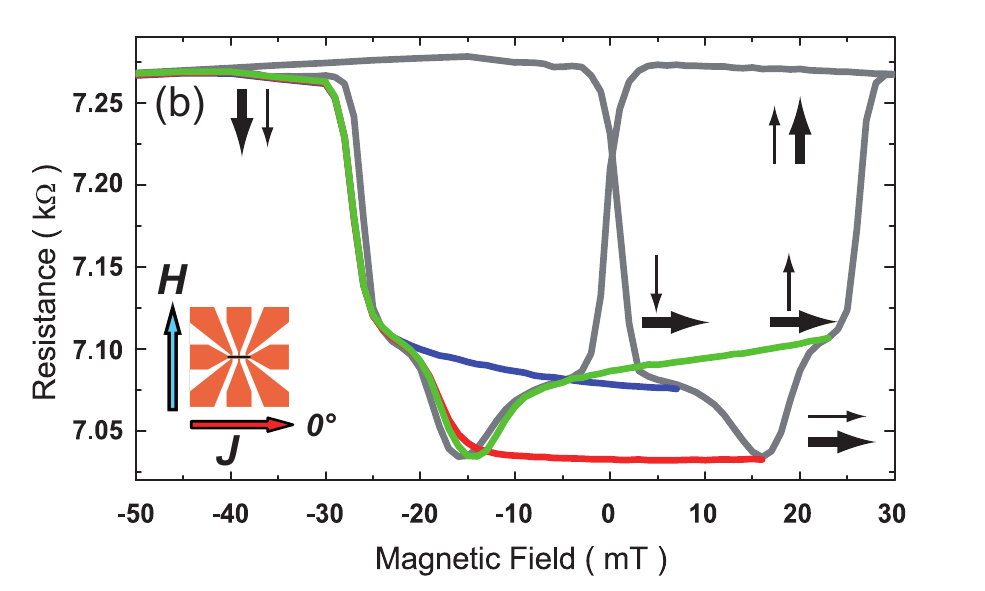}
\caption{a) MR scan at 4.2 K for a (Ga,Mn)As layer with 38 nm thick
stripes for various angles $\varphi$ between magnetic field and
current. The field is swept from -300 mT to 300 mT. b) MR scan at
$\varphi$ = 90$^{\circ}$ (light gray) and minor hysteresis loops
from -300 mT to the states at 7 (blue), 16 mT (red) and 23 mT
(green). The thick (thin) arrows indicate the magnetization of the
striped (bulk) regions of the sample at various positions in the
field sweep.}
    \label{fig:40nm_multi}
\vspace{-0.5cm}
\end{figure}

We now consider this transition from biaxial to uniaxial behavior in
a sample with stripe thickness of 38 nm. Fig. 3a shows AMR
measurements of such a sample. The high field behavior is again
unchanged. At low magnetic fields however, when $\vec{H}$ is swept
in directions nearly perpendicular to $\vec{J}$, the number of
switching events increases from the two distinct features observed
in Fig. 2, to four distinct features. Near $H$ = 0 mT a first
switching of $\vec{M}$ resulting in an intermediate resistance
occurs. A second switching event then follows decreasing the
resistance to its minimum value corresponding to $\vec{M}$ parallel
to $\vec{J}$. As $\vec{H}$ continues to increase the resistance
changes again to an intermediate level before returning to the
maximum resistance. Comparing these curves to those of
Fig.~\ref{fig:20nm60nm} leads to an intuitive interpretation of what
is happening: The sample breaks up into two regions having either
biaxial or uniaxial magnetic anisotropy.

Consider the gray curve of Fig. 3b showing an enlarged view of the
low field part of the $\varphi$ = 90$^{\circ}$ curve, and indicating
by the thick (thin) arrows the direction of magnetization of the
stripes (between the stripes) regions of the sample at various point
in the field sweep. At high fields, and down to $\sim$ -30 mT, the
field aligns all moments into a uniform state $\vec{H}
\parallel \vec{M} \bot \vec{J}$. As the field is reduced, a first
magnetization reorientation event occurs very near to $\vec{H}$ = 0
mT, as the magnetization of the uniaxial stripes changes to an
alignment along their easy axis, and thus parallel to the current.
This leaves the sample in a non-trivial configuration, where the
magnetization of the stripes and the regions between the stripes are
orthogonal to each other. As the field is further increased to about
16 mT, the magnetization of the biaxial regions undergoes a
90$^{\circ}$ reorientation as part of its two step switching
process. This reestablishes a co-linear, and thus low resistance
state. Further sweeping of the field towards high positive value
causes first the region between the stripes, and then the stripes to
switch their alignment to the magnetic field direction, creating a
second orthogonal state at 23 mT, and then the high resistance state
where all moments are perpendicular to the current, from +30 mT
onwards.

In order to verify that all these states are non-volatile, we
proceed to a series of minor loops to verify their hysteretic
behavior. The blue curve is obtained by sweeping the $\vec{H}$-field
from -300 to 7 mT, just after the first sharp switching event, and
then sweeping back to to negative fields. The sample clearly remains
in the intermediate resistance state confirming that the orthogonal
configuration of the magnetization in the different regions survives
in the absence of magnetic field. The same is true for the second
orthogonal configuration as well as the collinear low resistance
state, as confirmed by the next two experiments which repeat the
above procedure, but reverse field at 16 and 23 mT, at the point of
emergence of each relevant state.

\begin{figure}
    \includegraphics[width = 1.0\columnwidth]{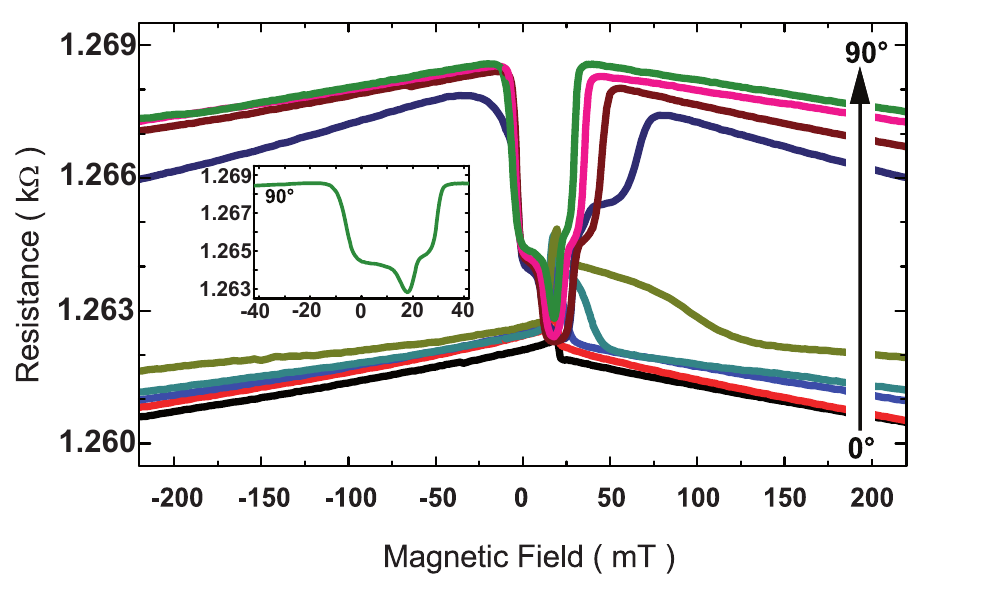}
\caption{Magnetoresistance measurements on the re-etched sample for
stripes of 37 nm thickness, showing the same mixed state behavior as
in Fig.~\ref{fig:40nm_multi}b. The inset shows the low field part of
the 90$^{\circ}$ curve.}
    \label{fig:S2_mixedstate}
\vspace{-0.5cm}
\end{figure}

To confirm the robustness and reproducibility of the results
presented, we have fabricated a second set of samples. The
fabrication is as above, except that the Ti mask is left on the
sample after processing. The sample can thus be measured for a given
stripe thickness and then re-etched in CAIBE, resulting in thicker
stripes that can be measured again. This allows us to investigate
stripes of various thicknesses on the same piece of material, and
thus ensure that the stripe thickness is the only parameter at work.
Figure~\ref{fig:S2_mixedstate} shows measurements on the re-etched
sample for a stripe thickness of 37 nm, and reproducing the mixed
state behavior of Fig.~\ref{fig:40nm_multi}b. In this case, because
the Ti mask which is still on the sample reduced the overall device
resistance, we use a four terminal measurement configuration to
fully exclude any role of the contacts.

\begin{figure}
    \includegraphics[width = 1.0\columnwidth]{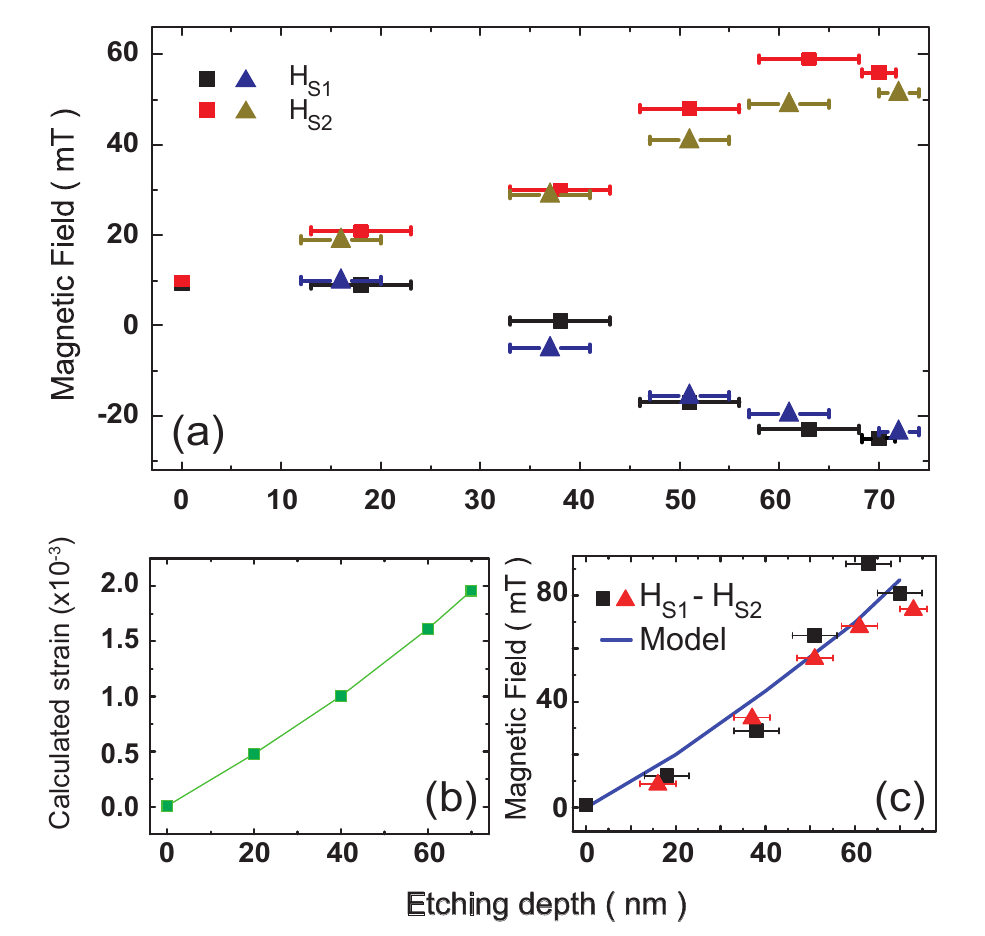}
\caption{a) Magnetic fields of the switching events
$\vec{H}_{{S}_{1}}$ and $\vec{H}_{{S}_{2}}$ for the individual
samples with different etching depths ($\blacksquare$) and the
subsequently etched sample ($\blacktriangle$). b) gives the average
strain $\epsilon_{x}$ as a function of the etching depth. c) plots
the separation between $\vec{H}_{{S}_{1}}$ and $\vec{H}_{{S}_{2}}$,
which is a measure of the uniaxial anisotropy strength. The solid
line is the anisotropy strength expected from modeling.}
    \label{fig:Hfields}
\vspace{-0.5cm}
\end{figure}

A figure of merit to quantify the strength of the imposed uniaxial
anisotropy term in the stripes can be obtained from the magnetic
field positions $\vec{H}_{{S}_{1}}$ and $\vec{H}_{{S}_{2}}$ where
the resistance reaches the mid value between its minimum and maximum
(see Fig.~\ref{fig:40nm_multi}a), for the two uniaxial switching
events in the $\vec{H} \perp \vec{J}$ curves of each sample
\cite{NatureParabola, Humpfner2007}. Results for both sets of
samples are plotted in Fig.~\ref{fig:Hfields}a, with the set
comprised of individual samples given as squares and the set from
re-etching as triangles. To within experimental determination of the
stripe thickness, both sets reveal quantitatively identical
behavior, and show a monotonic increase of the strength of the
imposed uniaxial anisotropy as a function of increased etching
depth. These results imply that the lithography is capable of
controlling the relative areas of biaxial and uniaxial regions and
allow for detailed engineering of anisotropic superstructures.

To better understand the role that strain relaxation plays in
setting up this mixed stated configurations, we turn to finite
element calculations. In Fig. 6a we present finite element strain
simulations on a samples with 40 nm thick stripes. No strain
relaxation occurs along the length of the strips, and we plot strain
components perpendicular to the stripe direction ($\epsilon_{x}$)
and ($\epsilon_{z}$). $\epsilon_{x}$ = 0.00 indicates full
pseudomorphic strain, where the lattice constant of the (Ga,Mn)As
layer is equal to that of the GaAs substrate. For $\epsilon_{x} > 0$
the (Ga,Mn)As lattice constant increases, indicating strain
relaxation, with $\epsilon = 1.96 \times$ 10$^{-3}$ corresponding to
the Ga$_{0.965}$Mn$_{0.035}$As fully relaxing to its natural lattice
constant. Negative values indicate stronger compressive strain than
that of the pseudomorphic layer. The simulations clearly show that
not only the stripes, but also the regions immediately under the
stripe are significantly relaxed in both $x$ and $z$ directions,
whereas the regions between the stripes are essentially
pseudomorphic, or even partially compressively strained. This
explains why the two regions have such different anisotropy
properties. Moreover, the fact that the relaxed strain extends to
include the material under the stripes explains why the uniaxial
region carries the majority of the current, as can be established
from the data of Figs.~\ref{fig:40nm_multi} and
\ref{fig:S2_mixedstate} where the larger of the two resistance
changes occurs when the stripes change their magnetization
direction.

The measurements presented in this manuscript are taken at 4.2 K. As
temperature is increased, the qualitative behavior remains unchanged
until about 20 K, after which the bulk layer undergoes the biaxial
to [110] uniaxial anisotropy transition commonly observed in
(Ga,Mn)As \cite{Sawicki2004}. The strength of the strain relaxation
induced uniaxial anisotropy in the stripes shows little temperature
dependence, as observed previously for this anisotropy mechanism
\cite{Humpfner2007}.

Finite element calculations were carried out on stripes of various
thickness, and the average strain in the striped region was
extracted for each thickness. These were used as input for $k\cdot
p$ calculations to extract the expected value of the anisotropy
strength. Following Ref.\cite{Abolfath2001,Dietl2001} we calculate
the ground state energy of the hole system under the assumption that
the magnetization is aligned along a given crystal direction. We
then repeat this calculation of all possible directions of
magnetization, to get a profile of the magnetic anisotropy, where
the directions yielding the lowest total energy are obviously
magnetically easy axes. For the strain Hamiltonian, we treat the
strain in the stripes as homogeneous and given by the average value
of the calculated strain in both the x and z directions (the y
direction is pseudomorphic) and the off diagonal components of the
strain matrix are taken to be zero. The average strain
$\epsilon_{x}$ as a function of etched depth extracted from the
finite element calculations, and used in the $k\cdot p$ modeling are
given in Fig. 5b. The other material parameters used in the modeling
are as in Ref. \cite{Schmidt2007}.

Using the methods described in Ref. \cite{Wenisch2007,Humpfner2007}
to relate the uniaxial anisotropy strength to the opening in the
magnetoresistance curves, we plot as the solid line in
Fig.~\ref{fig:Hfields}c, the expected opening from the model, and
compare it to the experimental values (solid symbols). Given the
oversimplification in the model that the strain for each thickness
can be reduced to a single homogeneous average value, the
correspondence between theory and experiment is quite remarkable.

\begin{figure}
    \subfigure{\label{fig:sim}}
    \includegraphics[width = 1.0\columnwidth]{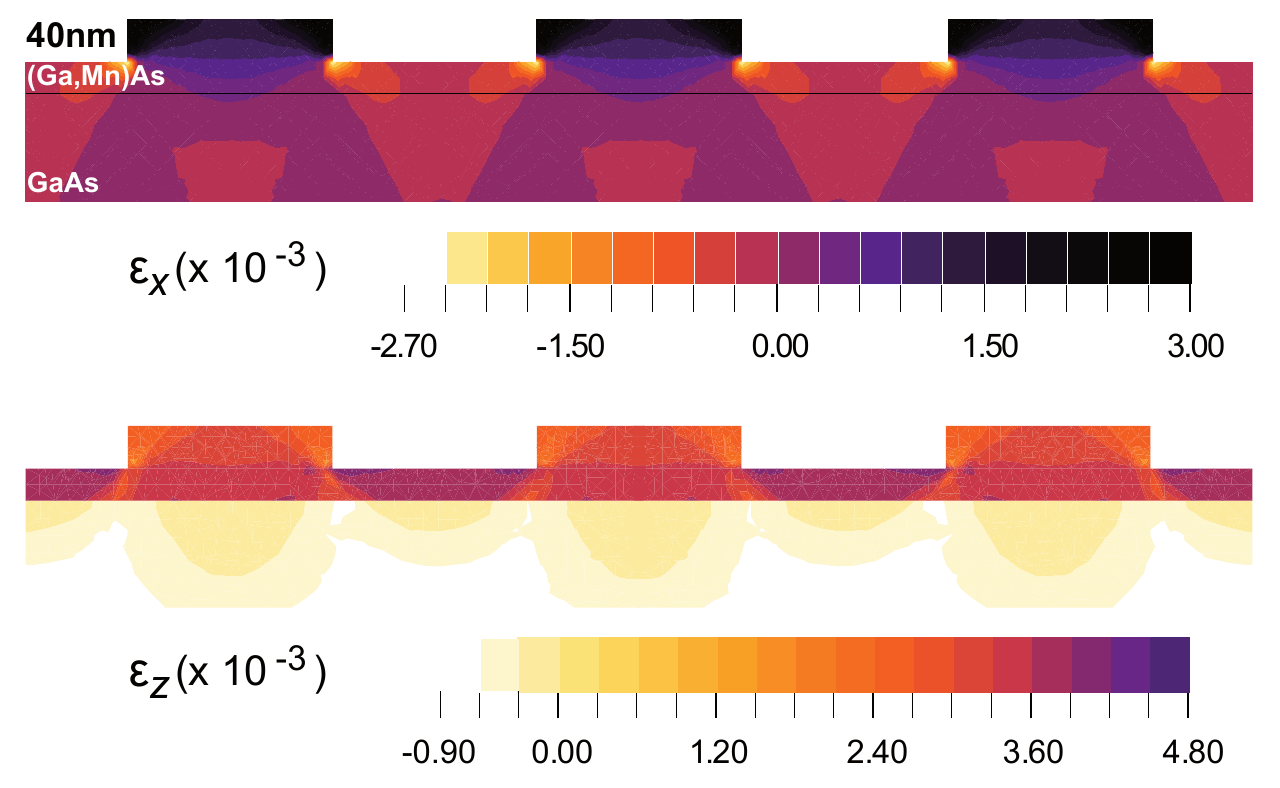}
\caption{Finite element strain simulations for stripes of 40 nm
thickness}
    \label{fig:simHfields}
\vspace{-0.5cm}
\end{figure}

In conclusion, we have demonstrated how carefully engineered strain
relaxation in (Ga,Mn)As can be used to create non trivial magnetic
anisotropies within a continuous layer out of which devices can be
patterned. We have used this method to demonstrate a new device
functionality. The structure behaves like a two state
magnetoresistance memory element, with a resistive response to the
difference between two non-volatile magnetization states. Unlike
previous devices where the two states differ in the relative
alignment of the magnetization of separate layers, here the two
regions of differing magnetization are contained within a single
layer. Apart from the obvious advantages in applications, this
method can also be useful for studies relying on the interplay
between regions of different magnetization, such as domain wall or
spin torque studies.

\begin{acknowledgments}
The authors thank T. Borzenko and V. Hock for assistance in sample
fabrication, and acknowledge financial support from the EU (NANOSPIN
FP6-IST-015728).

\end{acknowledgments}

\bibliographystyle{prsty}
\bibliography{controlpaper}

\end{document}